\documentclass[a4paper,11pt]{article}

\usepackage{jinstpub} 
 \pdfoutput=1
\usepackage{lscape}
\usepackage{lineno}
\usepackage{subcaption}
\usepackage{tablefootnote}
 \usepackage{hyperref} 
 \usepackage{url}
\usepackage{booktabs, caption, makecell}
\usepackage{threeparttable}

\title{\boldmath Positron Sources for Future High Energy Physics Colliders}

\bigskip 

\author[1]{P.~Musumeci}

\author[2]{C.~Boffo}
\author[3]{S. S.~Bulanov}
\author[4]{I.~Chaikovska}
\author[4]{A.~Faus Golfe}
\author[5]{S.~Gessner}
\author[6]{J.~Grames}
\author[7]{R.~Hessami}
\author[8]{Y.~Ivanyushenkov}
\author[9]{A.~Lankford}
\author[10]{G.~Loisch}
\author[10,11]{G.~Moortgat-Pick}
\author[2,12]{S.~Nagaitsev}
\author[10]{S.~Riemann}
\author[13]{P.~Sievers}
\author[14]{C.~Tenholt}
\author[15]{K.~Yokoya}

\affiliation[1]{UCLA Department of Physics and Astronomy, Los Angeles, CA 90095}
\affiliation[2]{Fermilab, Batavia, IL 60510}
\affiliation[3]{Lawrence Berkeley National Laboratory, Berkeley, California 94720, USA}
\affiliation[4]{Université Paris-Saclay, CNRS/IN2P3, IJCLab, 91405 Orsay, France}
\affiliation[5]{SLAC National Accelerator Laboratory, Menlo Park, CA 94025}
\affiliation[6]{Jefferson Lab, Newport News, VA 23606}
\affiliation[7]{Stanford University, Stanford, CA 94036}
\affiliation[8]{Argonne National Laboratory, Lemont, IL 60439}
\affiliation[9]{Department of Physics and Astronomy, University of California, Irvine, CA}
\affiliation[10]{Deutsches Elektronen-Synchrotron DESY, D-22607 Hamburg, Germany}
\affiliation[11]{University of Hamburg, D-22761 Hamburg, Germany}
\affiliation[12]{The University of Chicago, Chicago, IL 60637}
\affiliation[13]{CERN, CH 1211 Geneva 23, Switzerland}
\affiliation[14]{Helmholtz-Zentrum Hereon, D-21494 Geesthacht, Germany}
\affiliation[15]{High Energy Accelerator Research Organization (KEK), Japan}

\abstract{ 
An unprecedented positron average current is required to fit the luminosity demands of future e+e- high energy physics colliders. In addition, in order to access precision-frontier physics, these machines require positron polarization to enable exploring the polarization dependence in many HEP processes cross sections, reducing backgrounds and extending the reach of chiral physics studies beyond the standard model. The ILC has a mature plan for the polarized positron source based on conversion in a thin target of circularly polarized gammas generated by passing the main high energy e-beam in a long superconducting helical undulator. Compact colliders  (CLIC, C3 and advanced accelerator-based concepts) adopt a simplified approach and currently do not plan to use polarized positrons in their baseline design, but could greatly benefit from the development of compact alternative solutions to polarized positron production. Increasing the positron current, the polarization purity and simplifying the engineering design are all opportunities where advances in accelerator technology have the potential to make a significant impact. This white-paper describes the current status of the field and provides R\&D short-term and long-term pathways for polarized positron sources. 
}


\begin{document}
\maketitle
\flushbottom


\section{Executive Summary} 
Positron sources are a critical element for current and future e+e- colliders as luminosity requirements push the performances of these sources well beyond the current state of the art. For example, the International Linear Collider (ILC) plans to use average positron currents of 30 $\mu$A, nearly two orders of magnitude larger than any other positron source ever realized \cite{ILC:TDR}. In addition, there is a clear demand for high polarization control of the positron beam in order to improve the effective luminosity, reduce the background, and extend the reach of searches for beyond-the-standard-model chiral physics\cite{moortgat2008polarized}. As discussed in the Snowmass Energy Frontier report on future linear colliders \cite{Snowmass:ILC}, polarization of both beams is really needed to reap the benefits of the spin-dependence in the collision cross-sections. 

Within the Snowmass process, the importance of this topic has been recognized, as well as the lack of a coherent effort in the US accelerator physics community to tackle the challenges associated with very high current production of polarized positrons. It is worth to note here that the positron source is one of the future collider components where a relatively small investment (compared to the development of the main linear accelerator) has the potential to yield significant gains in the performances of the machine. In addition, the stand-alone nature of the positron source allows for tests and parallel developments that can be carried out in an independent fashion with respect to the main collider complex.

The generation of polarized positrons has been included in the baseline design of the ILC \cite{Yokoya2018, ilc-pos}. The scheme is based on passing the 125 GeV collision beam through a very long short-period superconducting helical undulator to generate circularly polarized gamma rays that can be converted using a thin target into  polarized electron-positron pairs. The ILC design is quite mature, at a level well beyond technical design report, and nearly shovel-ready.  Still, the reduction of the ILC center-of-mass energy to 250 GeV implies a lower-than-ideal gamma photon energy of 7-8 MeV from the undulator, at the falling edge of the pair production cross-section. To compensate for this, an extended undulator length (up to 231~m) will be employed to preserve a safety margin ($>$ 1.5) in the ratio of output positrons to incoming electrons. This effort shares many commonalities with existing activities in superconducting helical undulator development and characterization for X-ray FELs which are carried out with DOE Basic Energy Science funding at several US National Labs (FNAL, ANL, and LBNL) \cite{gourlay2016us,ivanyushenkov2015development}. 

The interdependency of the polarized positron source on the availability of the 125~GeV electron beam has also spurred a parallel effort in the development of a conventional positron sources based on a 6~GeV electron beam \cite{Nagoshi2020nima, resultsILCconv}. This conventional high current source will simplify the commissioning phase of the accelerator providing a reliable source of positrons. It will also allow to develop and test technical solutions for the challenges associated with the energy deposition in the target and the positron capture section immediately downstream of it. In particular, the target suffers from extremely large heat deposition rates, exacerbated in the polarized positron production case by the small transverse spot of the gamma-ray beam from the undulator and by the burst temporal format of the drive beam. The design for the ILC baseline source includes a rapidly spinning wheel with state-of-the-art ferrofluidic seals to allow for the fast rotation. 

The capture section after the target must be able to match the very large phase space of the emitted positrons into the small acceptance of the booster linac and finally into the damping ring. Quarter-wave transformers, flux concentrators and pulsed solenoids are typical solutions right after the conversion target. Simulations show that the very large magnetic fields achievable with a pulsed solenoid (currently the favored choice for the ILC) can increase the positron capture rate by 30 $\%$ with respect to the previously adopted quarter-wave-transformer scheme \cite{ushakov2018_matching}. On the longer term, the use of advanced solutions based on strong focusing lenses such as active plasma lenses has been considered and deserves further investigation\cite{formela2021designing}. 

In parallel to these efforts, other linear collider designs such as CLIC \cite{CLIC:CDR} and CCC \cite{nanni:ccc} do not foresee in their baseline the use of polarized positrons and linear collider schemes based on advanced accelerator concepts are just now beginning to consider the issues related to the accelerator of positrons. Improving the polarization purity and providing overhead in the positron average current, are all important goals in the development of future polarized positron sources. Two particular schemes are being considered based on Inverse Compton Scattering (ICS) and polarized bremmsstrahlung.

Polarized positron sources based on the Inverse Compton Scattering laser photons off energetic electrons have been proposed for a long time \cite{omori2003design} and are recently making a comeback fueled by the progress in laser technology. For example, scattering a circularly polarized 515~nm laser off 1~GeV electrons yields very energetic (30-40 MeV) polarized gamma rays. In this case, the yield of polarized positrons per incident photon can be up 3-5 times larger than when using $<$10 MeV photons. The scheme is plagued by the poor cross section of the Compton scattering process, but as GeV electrons are easily available, it is possible to increase the electron current to recover the required photon flux. One of the main outstanding problem of this scheme is the availability of high average power laser beams, but continuous progress in laser technology (for example laser stacking cavities \cite{akagi2012}) and new ideas in high efficiency free-electron lasers \cite{duris:tesso} open the opportunity for a compact, independent, polarized positron source with high flux and high polarization purity for future collider design.   

Polarized bremmsstrahlung is the process through which spin-polarized relativistic electrons (typically produced from strained GaAs lattice with very high ($>$80 \% polarization purity) can generate polarized positrons. Polarization transfer up to 80 \% have been demonstrated in a recent successful experiment at JLAB \cite{PhysRevLett.116.214801}, but the efficiency (number of positrons produced per incoming electrons) of this process is very small. Still, a collaboration has been formed around the idea of using this approach to generate polarized positrons for nuclear physics experiments. The positron currents from this source (50-100 nA) are still orders of magnitude lower than what required for linear colliders. Even lower positron production rates can be obtained using high intensity laser plasma interactions or isotope decays \cite{hessami:2022,li.prl.2020po}. These sources could possibly be used to provide a positron beam to test some of the charge-asymmetries in the high gradient accelerator schemes. 

\section{Positron sources for High Energy Physics colliders: requirements and current status}
\label{sec:intro}

\begin{figure}[htbp]
\centering 
\includegraphics[width=\textwidth]{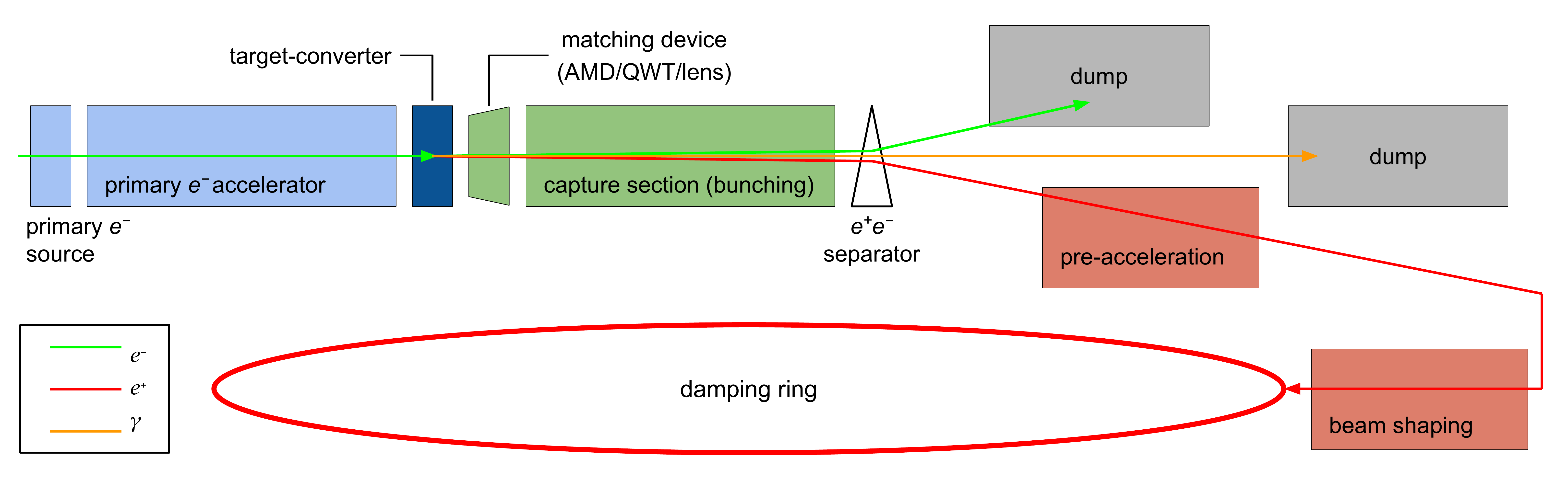}
\caption{\label{fig:main_layout} Different sub-systems of the positron source basic scheme.}
\end{figure}

In all positron sources used for high energy physics, positrons are produced by the process of pair production as as secondary beams after a drive beam (typically electrons, but it could be gamma photons) hits a conversion target. The resulting positron distribution has very large angular and energy spread and is captured transversely and longitudinally to match into an acceleration section and ultimately a damping ring to generate positrons beam of sufficiently high quality for the intended application. A cartoon schematics of the various elements of a positron source is shown in Fig. \ref{fig:main_layout}.

State-of-the-art positron sources parameters are summarized in Table \ref{tab:pos_source_perf} which shows that typical positron flux obtained is around 10$^{10}$ e+/s, several orders of magnitude lower than what required for the linar collider. The main limit in conventional sources is the heat load on the target which limits the power of the primary beam.  The SuperKEKB positron source is the highest intensity positron source in operation, thanks to improvements on the drive beam parameters, the flux concentrator used to capture the beam and improvements in the positron line diagnostics \cite{suwada2021first, enomoto2021}.

 \begin{table}[ht!]
 \caption{Performances of ever existed positron sources (readapted from \cite{chaikovska2022positron}. Some parameters were not found in literature and, therefore, marked as "--". List of the abbreviations used: Adiabatic Matching Device (AMD), Flux Concentrator (FC), Solenoid(Sol.), Quarter Wave Transformer (QWT), Linac End~(LE), Damping Ring (DR). The positron flux is calculated from the values listed in the table.}
\label{tab:pos_source_perf} 
\centering
\begin{tabular}{lcccccc}
\hline 
{\bf Facility} & {\bf SLC } & {\bf SuperKEKB} & {\bf DAFNE } & {\bf BEPCII}& {\bf LIL } \\ 
\hline 
{\bf Research center} & {\bf SLAC} & {\bf KEK} & {\bf LNF} & {\bf IHEP}& {\bf CERN } \\ 
\hline 
Repetition frequency, Hz & 120 & 50 & 50 & 50 & 100 \\
Primary beam energy, GeV & 30-33 &3.5 &0.19 &0.21 &0.2  \\
Number of $e^-$ per bunch &$5 \times10^{10}$ &$6.25 \times10^{10}$ &$\sim 1 \times10^{10}$ &$5.4 \times10^{9}$ &$2 \times10^{11}$\\
Number of $e^-$ bunches /pulse & 1 & 2 & 1 & 1 & 1  \\
Incident $e^-$ beam size, mm & 0.6 & $\sim$0.5  & 1 &1.5 & $\sim$0.5 \\ 
Target material &W-26Re &W &W-26Re &W &W   \\
Target motion & Moving & Fixed & Fixed & Fixed & Fixed \\
Target thickness/size, mm & 20, r=32 & 14, r=2 & - & 8, r=5 & 7, r= 8  \\ 
Matching device &AMD (FC) &AMD (FC) &AMD (FC) &AMD (FC) &QWT \\
Matching device field, T &5.5 &3.5 &5 &4.5 &0.83  \\
Field in solenoid, T &0.5 &0.4 &0.5 &0.5 &0.36  \\
Capture section RF band & S-band& S-band& S-band& S-band& S-band \\
${e^+}$ yield, $N_{e^+}/N_{e^-}$ &0.8-1.2 (@DR) &0.4 (@DR) &0.012(@LE) &0.015(@LE)  &0.006 (@DR)  \\
${e^+}$ yield, $N_{e^+}/(N_{e^-} E)$ 1/GeV &0.036 &0.114 &0.063 &0.073 &0.030 \\
Positron flux\tnote{*}, $e^+$/s &$\sim 6\times10^{12}$ & $2.5 \times10^{12}$ &$\sim 1 \times10^{10}$ &$4.1 \times10^{9}$ &$1.2 \times10^{11}$ \\
Damping Ring energy, GeV &1.19 &1.1 & 0.510 & No  & 0.5 \\
DR energy acceptance $\frac{\Delta E}{E}$, \% & $\pm$1 &$\pm$1.5 &$\pm$1.5 & No&$\pm$1 \\

References & \cite{slc-target, clendenin1996compendium} & \cite{prcom1} & \cite{dafne-linac, dafne-linac2}  &\cite{bepc-pos} & \cite{LEP_90, LIL, LIL-robert} \\
\hline
\end{tabular}
\end{table}

The requirement for positron polarization adds additional degrees of complexity in the design of the source, but directly stems from the physics demands for future colliders \cite{moortgat2008polarized}. Having simultaneously polarized $e^-$ and $e^+$ beams in fact, is a very effective tool for direct as well as indirect searches for new physics. Polarized beams offer new powerful analyses, provide added-value and optimize ---together with the clean and precise environment--- the physics potential of an $e^+e^-$ linear collider. The use of both beams polarized compared with the configuration with only polarized electrons can lead to an important gain in statistics and luminosity, reducing the required running time and increasing the search reach of the linear collider. In addition, it will allow to  decrease ---due to pure error propagation--- the polarization uncertainty originating from the polarimeter. The gain in the polarization accuracy is directly transferred to, for instance, the accuracy for the left-right asymmetry measurement and is therefore decisive for getting systematic uncertainties under control. Furthermore, the use of both beams polarized are important to identify  independently and unambiguously the chiral structures of interactions in various processes, several of these tests are not possible with polarized electrons alone, see \cite{Moortgat-Pick:2005jsx, ,Karl:2017xra,Karl:2019hes} and references therein. Simultaneously polarized $e^{\pm}$ beams offer new and more observable, e.g.\ double-polarized asymmetries, and allow to exploit even transversely-polarized beams that are powerful for detecting new kind of interactions ( e.g. tensor-like) or new sources of CP-violation. Already at the first energy stage of $\sqrt{s}=250$~GeV, the availability of polarized positrons would be essential to keep systematics under control, save running time and to match the precision promises~\cite{Fujii:2018mli}.


\subsection{Current plans for ILC positron source}

The positron production scheme for the high-energy linear $e^-e^+$-collider has been at the center of much debate with the result that two parallel approaches are currently being pursued: i) a baseline scheme which is based on passing the high energy electron beam through a helical undulator to generate an intense photon beam for the positron production in a thin target, and ii) a scheme based on the use of a separate and independent electron beamline to create (unpolarized) $e^+e^-$-pairs in a thick target.  

The efficiency of positron production in a conversion target together with the capture acceleration of the positrons is low, so in both cases it is a challenge to generate  the  $1.3\times 10^{14}$ positrons per second  that are required at the ILC collision point  (nominal luminosity).  However, using a helical undulator allows to produce a circularly polarized photon beam enabling the generation of a longitudinally polarized positron beam which is the reason it has been selected as the baseline option for the ILC ~\cite{TDR1,TDR31,TDR32}.

The high number of required positrons at a linear collider pushes the drive beam intensity up and causes a high thermal load on the target load. The target wheel has to be cooled as well as rotated at an appropriate speed in order to distribute the load sufficiently. The material must stand the cyclic load at elevated temperatures. Experimental tests were performed with the electron beam of the microtron in Mainz (MAMI) to simulate the cyclic load  as expected during ILC operation~\cite{Ushakov:2017dha,Heil:2017ump}. The results of the irradiation tests at MAMI and detailed simulation studies with ANSYS showed that the expected load at the ILC positron target is below the material limits~\cite{Dietrich:2019rts}. The radiated targets have been analyzed both via laser scanning methods as well as synchrotron X-ray diffraction methods, demonstrating that the ILC target will stand the load. The design include detailed plans for radiation cooling as well as rotating the target, see below. Detailed engineering solutions on theses issues are still outstanding, however no technical showstopper is anticipated here.

\section{Undulators for polarized positron production}

\begin{table}[ht!]
\caption{Parameters of ILC Polarized Positron Source Undulator\label{table:ILC_Undulator}.}
\centering
\begin{tabular}{lcccc} \hline
Parameter & units & 250 GeV & 350 GeV & 500 GeV \\ \hline
Electron beam energy ($e^+$ prod.) & GeV   & 126.5 & 178     & 253 \\ 
Undulator length & m     & 231 & 147     & 147 \\ 
Undulator period length & cm    & 1.15 & 1.15    & 1.15 \\ 
Undulator K &       & 0.92 & 0.75  & 0.45 \\ \hline
\end{tabular}
\end{table}

The helical undulator is one of the main components of an undulator-based polarized positron source. Parameters of the ILC polarized positron source helical undulator are given below in Table \ref{table:ILC_Undulator}. We refer to \cite{riemann2020updated} and references therein for updates since the Technical Design Report.

The required period length is as short as 11.5 mm which sets the challenge of fabricating the long helical undulator with such period length. This has been addressed and successfully solved by the UK HeLICal Collaboration. After an intensive R\&D phase~\cite{Ivanyushenkov_2005} , the collaboration has eventually fabricated and tested a $NbTi$ superconducting helical undulator prototype which has achieved the required parameters~\cite{Scott2011Undulator}, Table \ref{table:ILC_Undul_Prot}.

\begin{table}[hb!]
\caption{Parameters of ILC Undulator Prototype. \label{table:ILC_Undul_Prot}}
\centering
\begin{tabular}{lcc} \hline
Parameter              & units & value  \\ \hline
Cryostat length         & m   & 4    \\ 
Magnet length           & m   & 2 x 1.74      \\ 
Period length           & mm  & 11.5   \\ 
Undulator field on axis & T   & 0.86  \\ 
Magnetic bore diameter  & mm  & 6.35    \\ 
Vacuum bore diameter    & mm  & 5.85    \\ \hline
\end{tabular}
\end{table}

Advantage of employing a superconducting magnet technology for building a short-period helical undulator has been demonstrated, and a helical superconducting undulator (HSCU) has become a baseline for the ILC positron source undulator. Here a helical magnetic field is generated by a pair of helical electromagnetic coils with the currents in the opposite directions which are wound on the same magnet former. Compared to an alternative approach of using permanent magnets for generating helical field, the HSCU offers a natural simplicity of winding helical coils combined with a high magnetic field. 
The HSCU field can be increased further when high-field superconductors like $Nb_{3}Sn$ are employed instead of $NbTi$. A team at the Advanced Photon Source of Argonne National Laboratory, US has recently experimentally demonstrated that in a planar SCU the field is increased by at least 20 \% over the $NbTi$~\cite{Kesgin_2021}. Also, development of HTS-type superconductors is currently a very dynamic field with a high potential of reaching undulator field exceeding the one of a  $Nb_{3}Sn$ undulator. This has been shown in a small test planar undulators wound with HTS tape starting in 2010 ~\cite{HTS_SCU_CB} and has now reached the current densities in the winding higher than in the $Nb_{3}Sn$ ~\cite{Kesgin_2017}. Application of $Nb_{3}Sn$ and HTS superconductors in short-period helical undulators is therefore a topic for future R\&D with potential for significant impact.

\section{Positron target technology}
\begin{figure}[ht!]
\center
 \includegraphics*[width=130mm]{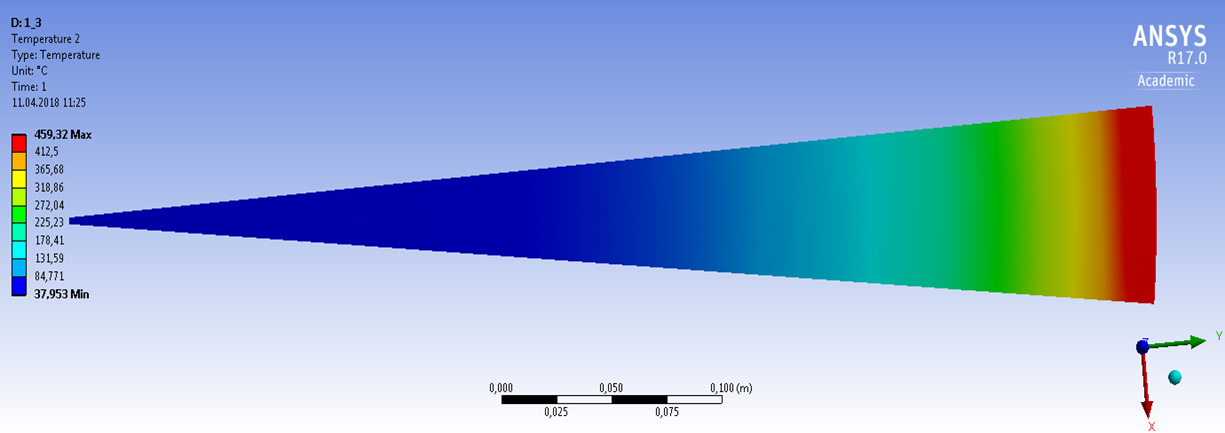}
  \caption{ Average temperature distribution in  the target shown for a sector corresponding to 1 pulse length (0.73\,ms) at ILC250; the beam impinges on the target at r=50~cm. The emissivities of target and cooler surface are 0.5 corresponding to an effective emissivity of $\varepsilon  = 0.33$.}   
 \label{fig:Tsector}
\end{figure}

The average energy deposition in the ILC positron target is about 2-7~kW depending on the drive beam energy in the undulator, the target thickness and the luminosity (nominal or high). As an example, for ILC250, the average energy deposition in the target is 2~kW. 
Energy deposition of up to few kW can be extracted by radiation cooling if the radiating surface is large enough and the heat diffuses fast enough from the area where the beam is incident to a large radiating surface.  In this design, the wheel spinning in vacuum radiates the heat to a stationary cooler opposite to the wheel surface. It is easy to keep the stationary cooler at room temperature by water cooling. But it is crucial for this scheme that the heat diffuses from the volume heated by the photon beam  to a larger surface area.  With the wheel rotation frequency of 2000rpm each part of the target rim is illuminated every 6-8 seconds, but this interval of time is not sufficient to distribute the heat load almost uniformly over a large area.  
The heat is then accumulated in the rim with the highest temperatures located in a relatively small region around the beam path. The average temperature distribution was calculated using the ANSYS software package~\cite{ansys} and is shown in figure~\ref{fig:Tsector} for one sector representative for the track of one bunch train.

For the studies of the positron yield optimization, the temperature distribution and  cooling principle  a target wheel designed as full 1~m-diameter disc of 7~mm thickness  made of Ti6Al4V was assumed. As expected, the radial steady state temperature in the wheel depends strongly on the radius. Due to the the heat conductivity in the target material and the $T^4$ dependence in the Stefan-Boltzmann law, most of the heat is removed close to the rim of the wheel. One should note, that by increasing the outer radius of the wheel up to 60 cm, while maintaining the beam impact at r=50~cm, substantially lower average temperatures can be expected. 

\begin{figure}[ht!]
\center
 \includegraphics*[width=130mm]{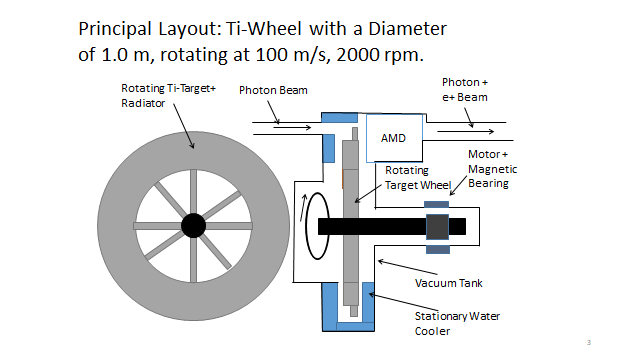}
  \caption{Principal Layout of the rotating wheel showing its main components: its cooling system, its rotating magnetic bearing and the  matching device (AMD). }
 \label{fig:peter-layout}
\end{figure}

Thus it is possible to  conceive a target wheel consisting of two distinct parts with separate functionalities: i) a 'carrier wheel', designed and optimized in terms of weight, material,  moment of inertia,  centrifugal forces, stresses and vibrations, etc., and ii) a second unit, the actual Ti-target rim. The target units are fitted mechanically to the rim of the carrier wheel in such a way that the cyclic loads, temperature rises and stresses in the target units are not or little transmitted to the carrier wheel.  This would allow to design and optimize the engineering of the carrier wheel independently from that of the target proper.  A possible layout in Figure~\ref{fig:peter-layout} shows the main items of the target wheel, the spoked rotating carrier wheel with its magnetic bearings and the water cooled stationary coolers~\cite{riemann2020updated}. 
Another interesting development in terms of target technology is the so called two-stage process for positron production. The first stage is optimized for the generation of photons/gamma rays (for example using channeling radiation in crystals). The charged particles in the EM shower get separated away using a magnetic field so that only the photons hit the second stage target improving the heat load and yield for a given drive beam intensity \cite{chaikovska2017_channeling, yoshida1998positronchanneling}. 

\section{Positron capture schemes for the ILC}


\subsection{Flux concentrator}
Most of these studies for positron capture after the target assumed a pulsed flux concentrator (FC) as optical matching device (OMD). A promising prototype study for the FC was performed by LLNL~\cite{ref:Gronberg-FC}. However, detailed studies identified some weak points in this design. The B-field distribution along $z$ cannot be kept stable over the long bunch train duration and therefore the luminosity would vary during the pulse which is not desired. Further, the particle shower downstream the target causes a high load at the inner part of the flux concentrator front side, which is at least  for ILC250 beyond the recommended material load level~\cite{ref:AU-OMDyield}. This is mainly caused by the larger opening angle of the photon beam and the wider distribution of the shower particles downstream the target at ILC250. As alternatives are discussed the use of a quarter-wave-transformer or a pulsed solenoid or --as an example for new technology-- a plasma lens.

\subsection{Pulsed solenoids}
Apart from the matching devices which are currently in use at positron sources in different facilities, like flux concentrators and quarter wave transformers, pulsed solenoid magnets have also been employed, e.g., at the LEP \cite{Warner1988}. %
Due to the limited yield that a quarter wave transformer can provide \cite{Yokoya2018}, interest in using a pulsed solenoid as an optical matching was renewed recently \cite{riemann2020updated}. \\%
To evaluate whether a pulsed solenoid would provide a sufficient yield, a stable magnetic field over 1~ms, and would not cause an excessive amount of heating in the rotating target wheel due to induced eddy currents, simulations have been performed in a collaboration involving CERN, University Hamburg, DESY, and the Helmholtz-Zentrum hereon. %
\begin{figure}[hb!]%
\centering
\includegraphics[width=0.5\columnwidth]{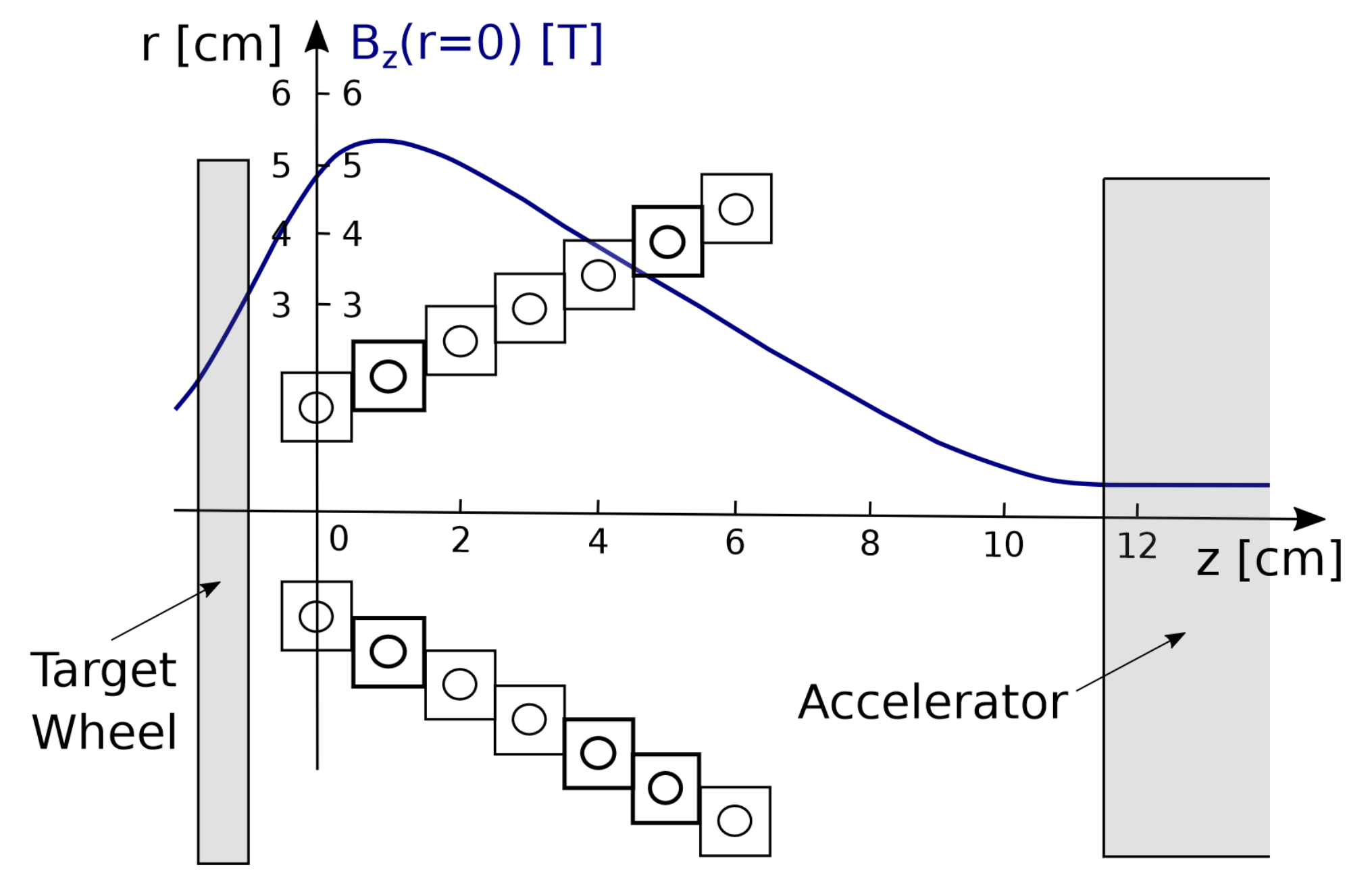}%
\caption{Sketch of the pulsed solenoid optical matching device for the ILC undulator-driven positron source \cite{riemann2020updated}.}%
\label{fig_PSsketch}%
\end{figure}
The principle layout is depicted in Fig.~\ref{fig_PSsketch}. %
A coil of 7 windings, with a tapered inner diameter of 20~mm at the target end and 80~mm at the downstream end, is formed by a square-shaped copper conductor with a circular inner cooling channel. %
The length of the quadrupole is 70~mm. %
According to simulations made in Comsol Multiphysics, a peak magnetic field of 5\,T is produced by applying a pulsed current with a peak amplitude of 50~kA. This field can be slightly increased by introducing a magnetic shielding made of ferrite around the solenoid. %
The field deviation over 1\,ms is found to be well below 1\,\% when applying a pulse with 2\,ms sinusoidal rise time, a flat-top current of 1\,ms duration and another sinusoidal fall time of 2\,ms. \\%
Using a ferrite shielding also reduces the magnetic field at the target position and therefore reduces heating in the target wheel due to eddy currents. %
This heating was also simulated and the expected values of the peak and average heat load, as well as the peak force on the target wheel are expected to be well manageable. %
Similarly, no critical values have been found for the thermal load and mechanical stress in the coil itself. \\%
The positron yield of an undulator-based positron source with a pulsed solenoid as a matching device was also simulated. Without ferrite shielding, a yield of 1.9 positrons per electron was simulated at the ILC positron damping ring. When using a ferrite shielding, the yield was slightly reduced to 1.7. For comparison, the positron yield using a quarter wave transformer, which is currently the baseline design option for the ILC, is only 1.1. Further increase of the yield with the pulsed solenoid might be possible by further optimisation of the exact coil geometry. In summary, pulsed solenoids are a viable option as a matching device for positron sources compared with current state-of-the-art solutions like quarter wave transformers and flux concentrators and especially for long bunch trains (as in the case of the ILC undulator-based positron source), simulations indicate that such a device would bring some advantages compare with the other options. %

\subsection{Plasma lens}

An alternative device to capture positrons after the target is an active plasma lens (APL) \cite{PanofskyBaker}. These focusing elements exhibit several advantages compared to conventional focusing elements like solenoids or quadrupoles:
\begin{itemize}
	\item due to the azimuthal magnetic field the focusing is radially symmetric unlike, e.g., in a quadrupole field
	\item focusing fields are potentially very high due to the close proximity of focusing currents and focused particles
	\item focusing fields are transverse to main direction of motion unlike, e.g., in a solenoid
	\item mitigation of space charge forces between beam particles due to the quasi-neutrality of the plasma medium
	\item low scattering of beam particles due to the low density of the conductive medium compared to, e.g., a Lithium lens or a magnetic horn.
\end{itemize}

In the particular application as a positron capture device, the APL has additional advantages over focusing schemes with solenoidal fields as flux compressors, solenoids or quarter wave transformers: fields are localised, i.e., do not influence the positron target wheel and the focusing is selective with respect to the particle charge that is when the active plasma lens is focusing positrons it is defocusing co-propagating electrons at the same time. These unwanted, low energy pair-electrons from the positron source will therefore not be accelerated in the capture linac, which reduces beam losses and radiation at high energies in the downstream accelerator significantly and also renders a dedicated charge separation chicane and a high energy electron beam dump unnecessary.

The usage of an active plasma lens as a matching device at a positron source was proposed for LEP \cite{Braunetal} and again recently for the ILC \cite{FormelaArxiv}. %
Especially due to the advances in development of high-gradient, beam quality preserving active plasma lenses in recent years \cite{vTilborgAPL,RoeckemannAPL,LindstromAPL,vTilborgFODO}, their application as focusing elements rather than as research objects on their own is now in reach and partially already the case \cite{SteinkeStaging}. Nevertheless, the positron source of the ILC poses several challenges for an active plasma lens to be used as an optical matching device (OMD) including the close proximity to accelerating cavities, which require ultra-high vacuum conditions, the large beam size (up to 1.5~mm) and very strong divergence of positrons and the challenging time format with short bunch separation of 554~ns in a train of $>$1000 bunches. On the other hand, beam-quality preservation is not a critical issue for an active plasma lens as an OMD for the low beam quality positron bunches at the source.

To investigate the possibilities for APLs to meet these challenges, a project has been initiated at the University Hamburg, Germany (UHH) and the Deutsches Elektronen-Synchrotron DESY in Hamburg. First results indicate that an APL indeed allows to increase the positron yield significantly w.r.t.\ the quarter wave transformer baseline design \cite{FormelaArxiv}. %
It should be noted though, that the simulated APL which allowed such an increased yield had a complex taper. %
Tapered lenses have been studied before \cite{TauschwitzTaper} but the available data is still very limited compared to more simple, linear discharge channels. %
Studies at the UHH and DESY are concentrating on investigating the field distribution within the APL in such a complex geometry and at high repetition rates both experimentally and in numerical simulations as well as on the question whether the yield improvement and required vacuum levels in nearby accelerating structures can be achieved at the same time.\\ %
Other groups are also looking into plasma lenses as capture optics for highly divergent beams at the source \cite{YangProtonAPL} and while the requirements of the ILC positron source e.g.\ in terms of repetition rate are certainly very demanding for state-of-the-art APL technology, plasma lenses can still be considered an option for other positron sources with different requirements in the future. %

\section{Novel approaches to polarized positrons}

\subsection{Compton-based polarized positron sources}
Another attractive and compact solution foresees the use of an high-power laser beam and the Inverse Compton Scattering (ICS) interaction to create such photons~\cite{bessonov1996method, okugi1996proposed}.
The electron beam requirements in this case are greatly reduced while still reaching higher photon energies. Considering the scattering with a typical laser ($\lambda$~=~515~nm), the electron energy required to generate 30~MeV photons is around 1.0~GeV and very small spot sizes have to be maintained only over relatively short interaction lengths (less than few cms). 
In 2005, a proof of principle experiment for the Compton scattering-based scheme for polarized positron generation was performed at the KEK Accelerator Test Facility (ATF)~\cite{PhysRevLett.96.114801}.

Several options for the future linear collider positron source based on
Compton scattering have been proposed~\cite{rinolfi2009clic}. Today, they can be classified according to the electron source used for the Compton scattering: the linac scheme, the storage ring scheme or so-called Compton Ring and the energy recovery linac scheme.
For all of them, the polarized positron current produced is not sufficient to fulfill the future linear collider requirements. Therefore, the application of the multiple-point collision line and multiple stacking of the positron bunches in the DR were investigated~\cite{chaikovska2012polarized, zimmermann2009stacking}. 

On the other hand, owing to the small size of the Compton (Thompson) cross section, the demands of such solution on the high-power laser system are extremely challenging\footnote{In case of Gamma Factory proposed at CERN~\cite{2019gamma}, which uses partially stripped ion beams and their resonant interactions with laser light, the resonant photon absorption cross section can be up to a factor $10^9$ higher than for the ICS of photon on point-like electrons. The proof of principle experiment was already proposed~\cite{Krasny:2690736}.}. 
The time format of ILC beams, for example, is constituted by an elevate number of bunches ($>$1000) per RF macropulse, with macropulse repetition rates of 5-10~Hz. At visible wavelengths, Joules of energy are required in order to provide sufficient photon density for the generation of one photon per incoming electron in the laser-beam interactions. The laser system should, therefore, provide multi-MW-class average power within a burst mode matching the electron bunch time-format. Using the additional degree of freedom offered by fast kickers, one can imagine to reformat the positron source to 30~KHz repetition rate and recreate the collider bunch format only after the DR, easing somewhat the peak and average requirements on the laser. Notwithstanding the exceptional progress of the RF and of the laser technology in the last decades, even this latter kind of laser system does not exist yet.
Various new concepts, such as stacking cavities and optical energy re-circulation have been proposed to address the lack of a suitable laser source for this application~\cite{chaikovska2016high, PhysRevAccelBeams.22.093501, PhysRevAccelBeams.23.051301, Amoudry:20, PhysRevAccelBeams.21.121601}.

In Murokh et al.~\cite{rbtoptical} the authors present
an alternative approach for an independent high-current polarized positron source based on the combination of laser-based acceleration with the observation, that the electron and laser beams are only minimally degraded in a ICS interaction. The laser pulse can then be used not only to drive the Compton scattering process, but also to accelerate the electrons to the required GeV-level for energetic polarized photon production. At the same time, after the ICS interaction point, the kinetic energy stored in the electrons can be recuperated with an high efficiency Free-Electron Laser (FEL) amplifier operating in the Tapering Enhanced Stimulated Superradiant Amplification (TESSA)  regime to replenish the laser pulse before it is redirected to scatter against the next electron bunch. Due to the limited electron beam and laser power requirements of this scheme, the electron current used in the accelerator can be very large and, even with the yield of 0.1~$N_{e^+}/N_{e^-}$ after conversion of the gamma-rays in the target, positron fluxes of up to 10$^{15}$ $e^+$/s could be achieved. 

It should be emphasized that due to a common technological constraint of all the above mentioned schemes being the average laser power of the optical systems, the Compton scattering-based polarized positron sources are considered only as the alternative solutions for the future collider projects.
Presently, it is proposed as a preferred option for an upgrade of the CLIC positron source~\cite{CLICcdr}.

\subsection{Polarized bremmsstrahlung}
Important topics in nuclear, hadronic, and electroweak physics including nuclear femtography,
meson and baryon spectroscopy, quarks and gluons in nuclei, precision tests of the standard model,
and dark sector searches may be explored at CEBAF, especially when considering potential upgrades in luminosity, polarized and unpolarized positron beams and doubling of the beam energy to 24 GeV~\cite{Arrington:2021alx,Accardi-2020}.

For a positron program, Polarized Electrons for Polarized Positrons (PEPPo) represents a pathway
to generate the highly spin-polarized positron beams required.  The technique is based upon the electro-magnetic shower of electron beams in matter and the two-step process of polarized bremsstrahlung followed by polarized pair creation~\cite{Bessonov-1996,POTYLITSIN1997395}.   Both steps can occur in a single high-Z conversion target, or be accomplished using a separate radiator and converter, if desired.  Notably, this technique can be applied at any electron accelerator where spin polarized electron beams are produced, whether at a university or national lab. 

\begin{figure}[htbp]
\centering 
\includegraphics[width=0.8\textwidth]{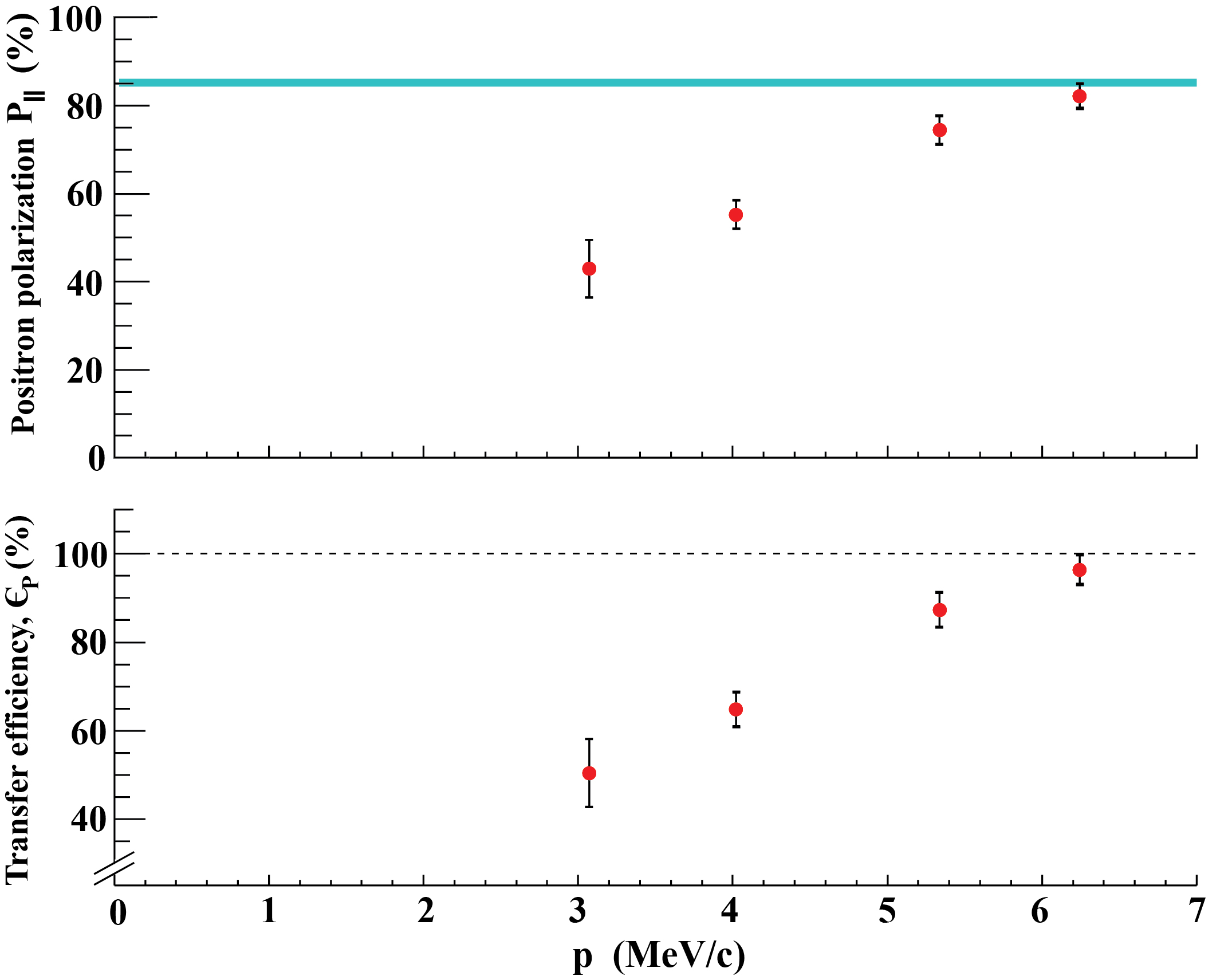}
\caption{\label{fig:peppo} Positron polarization and transfer efficiency as reported in ~\cite{PhysRevLett.116.214801} The blue line represents the electron beam polarization.}
\end{figure}

The transfer efficiency of spin polarization from the electron beam to the positron beam, defined as $\epsilon = P(e+)/P(e-)$, can be very efficient, approaching unity as the momenta of the collected positrons approaches the initial electron beam momentum. The technique was first demonstrated at CEBAF~\cite{PhysRevLett.116.214801} where an 8.2 MeV/c electron beam with polarization 85.2\% produced positrons with polarization >82\% (see Fig.~\ref{fig:peppo}).  Collecting the positrons at half the electron beam momentum serves to maximize figure of merit defined as IP$^{2}$, with positrons receiving >60\% of the electron beam polarization. 

In contrast to positron polarization, the positron yield N(e+)/N(e-) falls precipitously with increasing positron momenta, due to the characteristic bremsstrahlung power spectrum.   While this is not a deciding factor for unpolarized positron sources which select a low-momentum fraction of positrons from the conversion target, a PEPPo-driven polarized positron source must select the high-momentum fraction to provide polarization. Limitations in electron spin polarization and beam intensity likely explain why a PEPPo-based positron source has not been constructed to date.  However, this landscape has changed significantly in the last 10 years. Electron beam polarization is now routinely $\approx$ 90\% and with average beam currents at milliAmpere level~\cite{Grames:PSTP2017,doi:10.1063/1.5040226}.

Today, strained-layer superlattice (SSL) photocathodes composed of quantum well multi-layer heterostructures provide very high spin polarization >85\% and with yields $\approx$ 6 mA/W of laser light.  And SSL photocathodes fabricated with an integrated diffracted Bragg reflector - to more efficiently absorb optical power - have demonstrated yields >30 mA/W~\cite{doi:10.1063/1.4972180}.  One may now reasonably imagine providing 100 kW of highly spin polarized electron beam at energies in the range of 10-100 MeV.  In this context, a recent Jefferson Lab LDRD project~\cite{GramesLDRD2021} explored the possibility of generating >100 nA positron beams with polarization >60\% for experiments at Jefferson Lab’s Hall B~\cite{CLAS:2021gwi}, using a 1 mA spin polarized electron beam at 100 MeV. The results were encouraging and are now under further consideration as a potential future upgrade for CEBAF.

In summary, PEPPo demonstrated a compact and efficient technique to produce highly spin polarized positrons, suitable for small to large-scale accelerator facilities. Advances in GaAs photocathodes capable of producing a high degree of spin polarization and milliAmpere intensities makes this technique viable. It is recommended to the P5 panel to support R\&D in the areas of high current polarized electron sources, ~100 kW high power targets and magnets for efficient collection of positrons over energies 10-100 MeV.

\subsection{High intensity laser-based positron polarization}
The positron production using high intensity lasers was extensively studied over the last two decades employing a number of different mechanisms and interaction setups mostly analytically and in computer simulations, though a number of experimental studies was also reported \cite{dipiazza.rmp.2012,gonoskov.rmp.2022,fedotov.arxiv.2022}. 

The most straightforward one is the interaction of a moderate intensity laser with a solid-density target, which is several millimeters thick \cite{chen.prl.2009,chen.prl.2015,liang.sr.2015}. Here, the electrons accelerated by the laser at the front surface go through the target emitting photons along the way due to the bremsstrahlung. These photons create electron-positron pairs in the course of their interaction with nuclei. In principle, a high-energy electron beam can be used instead of the laser pulse in a positron production scheme, which is typical for conventional positron sources. In such scheme a high intensity laser is used to accelerate electrons via laser-wakefield acceleration \cite{gahn.apl.2000,sarri.prl.2013,sarri.ncomm.2015,roadmap.doe.2016,alegro.arxiv.2019,alegro.arxiv.2020}.  

\begin{figure}
    \includegraphics[width=0.8\linewidth]{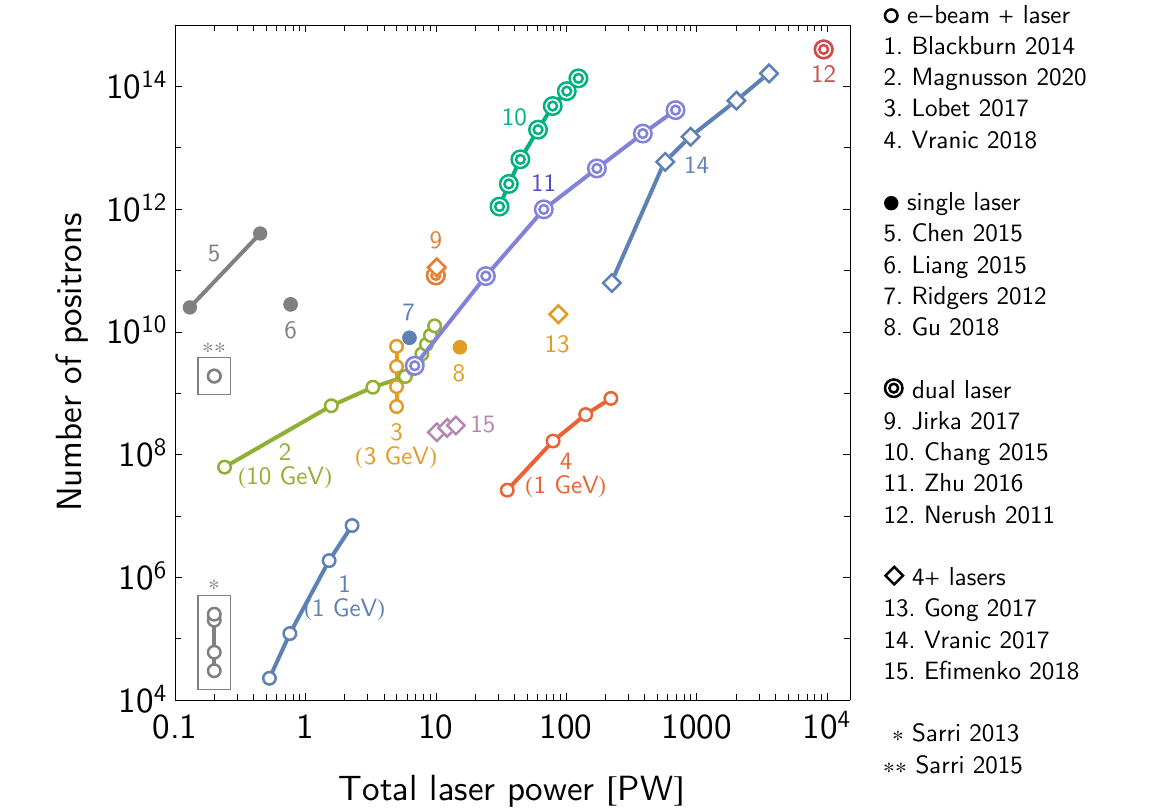}
    \caption{%
        Number of positrons produced in high-intensity laser-plasma interactions.
        For laser-electron beam interactions (open circles), the energy of the electron beam is noted in brackets.
        Points marked with asterisks indicate experimental results from LWFA electron-beam interactions with high-Z foils \cite{sarri.prl.2013,sarri.ncomm.2015}; in these cases the laser power is not indicated. Reproduced from \cite{gonoskov.rmp.2022}.
    }
    \label{fig:PositronYield}
    \end{figure}

The positron production using high energy lasers as converters of high energy photons into electron-positron pairs is based on the effects of strong field quantum electrodynamics (SFQED)  \cite{dipiazza.rmp.2012,gonoskov.rmp.2022,fedotov.arxiv.2022}. Here an electron beam interacts with a single high intensity laser pulse or a combination of several pulses, or instead of an electron beam a fixed plasma target is used \cite{ridgers.prl.2012,gu.cp.2018,jirka.sr.2017,chang.pre.2015,zhu.ncomm.2016,nerush.prl.2011,gong.pre.2017,vranic.ppcf.2017,efimenko.sr.2018}: high energy gammas are emitted by electrons going through a region of strong EM field via multi-photon Compton process, and these gammas decay into electron-positron pairs via the multi-photon Breit-Wheeler process (see Fig. \ref{fig:PositronYield}). We note that the production of electron-positron pairs is very sensitive to the EM field strength: for three orders of laser intensity the number of positrons varies ten orders of magnitude. There is an advantages of using a high energy electron beam interaction with a high intensity laser pulse, the positrons are produced as a collimated beam and the required laser intensity is much lower \cite{blackburn.prl.2014,lobet.prab.2017,vranic.sr.2018,magnusson.prl.2019}. 

The use of polarized electron beams in the above mentioned schemes will, first, lead to the production of polarized  $\gamma$-ray beams \cite{li.prl.2020po} and, subsequently, to the production of polarized positron beams. It is due to the fact that multi-photon Compton and Breit-Wheeler processes depend on the spin of participating particles. However, most of the reported studies use initially unpolarized electron beam and rely on its polarization during the interaction with a high intensity laser, which needs to be shaped in a way that breaks the symmetry of field oscillation to achieve net polarization. This can be achieved with  a two-color laser pulse~\cite{song.pra.2019,seipt.pra.2019} or with a laser pulse  with a small degree of ellipticity \cite{wan.plb.2020}. For example,  an initially unpolarized 2-GeV electron beam interacting with  a two-color laser pulse, with $a_0 = 83$ and 25\% of its energy in the second harmonic \cite{chen.prl.2019} acquires an average polarization degree of only 8\%, whereas the positrons produced have a polarization degree of 60\% due to the Breit-Wheeler process depending more strongly on spin than the Compton one.  A laser pulse  with a small degree of ellipticity can, in principle, generate positron beams with a polarization degree exceeding 80\%~\cite{wan.plb.2020}. 

In summary, it was theoretically shown that the polarized positron production using high intensity lasers can be achieved, however, the characterization of the phase space of these positron beams need to be carried out in future studies, as well as the study of their capture by beam transport systems and subsequent injection into an accelerator. The proof-of-principle experiments are required to access the possibility of using such positron source for compact colliders  (CLIC, C3 and advanced accelerator-based concepts \cite{benedetti.arxiv.2022a,benedetti.arxiv.2022b}). 

\subsection{Electrostatic traps as a test-bed for polarized positron physics}

The generation of positron beams is an expensive process requiring significant infrastructure. Experimental tests on positron beams are limited to facilities that are already equipped with a high-energy, high-intensity electron beam accelerator, a high-power target, and damping ring for cooling. As a result, very few institutions provide access to positron beams for experimental use. An alternative, compact system for producing polarized positron beams could provide experimental opportunities for testing systems associated with positron beam production and transport.

\begin{figure} [!htb]
\centering
\begin{subfigure}{.5\textwidth}
  \begin{center}  
  \includegraphics[width=0.9\linewidth]{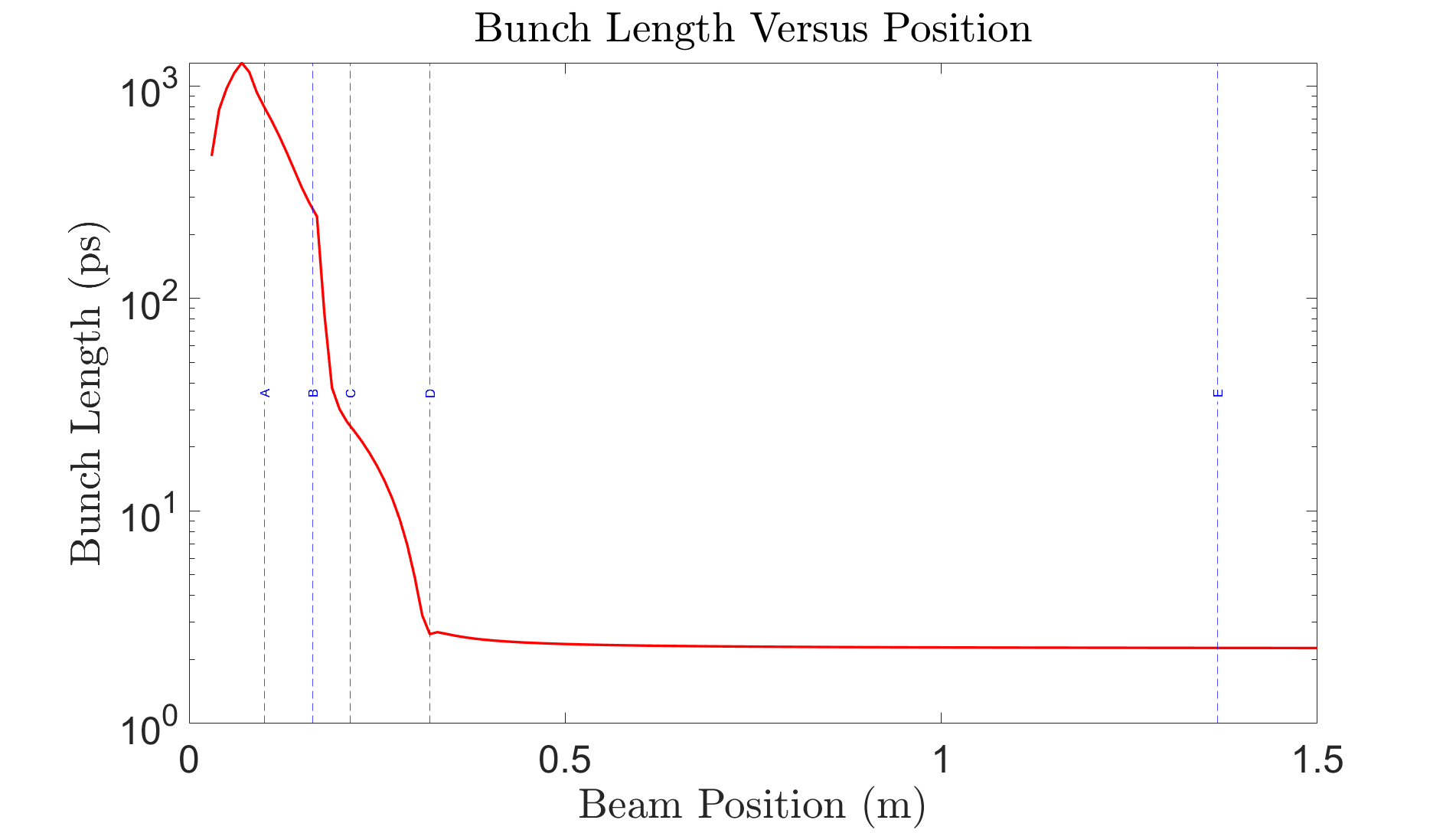}
  \end{center}  
\end{subfigure}%
\begin{subfigure}{.5\textwidth}
  \begin{center}  
  \includegraphics[width=0.9\linewidth]{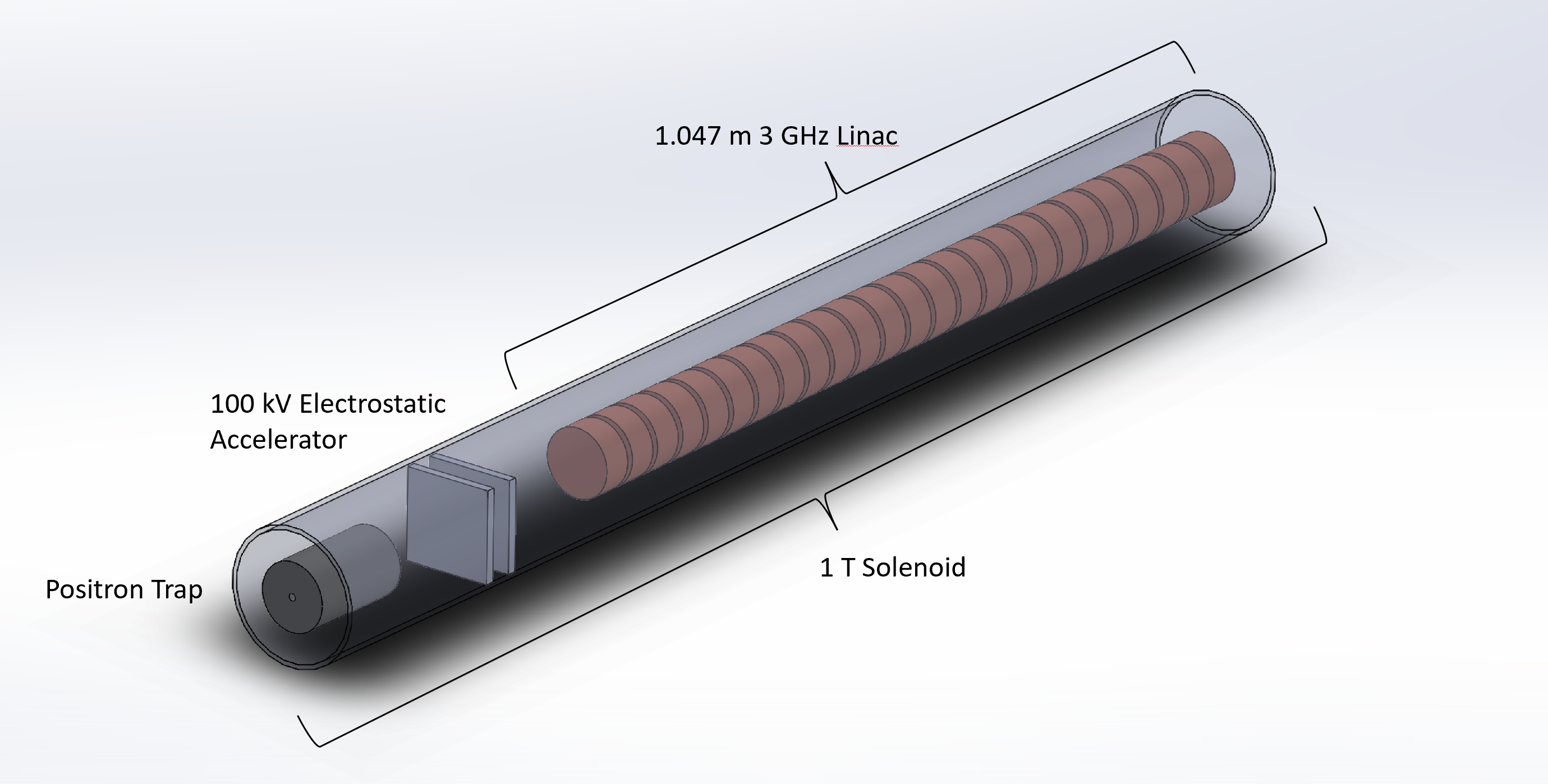}
  \end{center} 
\end{subfigure}
\caption{a) Positron bunch length along the beamline. b) Depiction the beamline design used in the simulation.}
\label{fig:positrontrapfigs}
\end{figure}

We propose a beamline design utilizing an electrostatic positron trap as a beam source for positron beams that is comparatively inexpensive and small~\cite{hessami:2022}. The concept is shown in Figure~\ref{fig:positrontrapfigs}. In this proposal, the positrons can be generated either by emission from a $\beta$-decay emitter, such as $^{22}$Na, which produces roughly $10^9$ positrons per second, or by impacting a 5 MeV electron beam on a high-Z target. The positrons pass through a solid-neon moderator which reduces their energy so that they can be trapped~\cite{doi:10.1063/1.97441}. The electrostatic trap holds the positrons while they accumulate and cool via interaction with a buffer gas. The longitudinal trap potential is shaped by high-voltage rings, and a solenoidal magnetic field provides radial confinement~\cite{PhysRevLett.85.1883,RevModPhys.87.247,GBARproposal}.

After positrons are accumulated in the trap, the trapping potential is changed to accelerate and eject the beam. The beam is both long and non-relativistic when ejected from the trap. The remainder of the beamline is dedicated to compressing and bringing the beam up to relativistic energies. To accomplish this, an electrostatic accelerator of 100 kV is employed, which compresses and accelerates the beam to the point that it can be injected into an s-band cavity. The beam is compressed to a bunch length of 0.2 mm and accelerated to an energy of 17.8 MeV. 

The entire beamline is inside of a 1 T solenoid. The beam is cooled inside a magnetic field and has intrinsic angular momemtum $\mathcal{L}$. The effective emittance is given by ~\cite{PhysRevSTAB.6.104002}:
\begin{equation} \label{angular_emit}
    \epsilon_n = \sqrt{\epsilon_{th}^2+\mathcal{L}^2}.
\end{equation}
With a small thermal emittance, the beam is dominated by angular momentum.


Future linear colliders assume that the emittance in the vertical plane is much smaller than in the horizontal plane because the beams are generated in a damping ring ~\cite{ILC2013TDRvol3Accelerator}. Our example beamline is capable of producing flat beams for ILC-type applications~\cite{Brinkmann2001}. While this compact source is not a suitable candidate for future Linear Colliders, it may be useful for testing positron capture technology or for demonstrating transport of flat beams.


\acknowledgments
The authors want to acknowledge all the participants to the Snowmass Polarized Positron Source workshop which was held by Zoom on March 1st 2022. The workshop agenda and presentations can be found at the following URL: https://indico.fnal.gov/event/52959/

\bibliographystyle{unsrt}
\bibliography{bibliography}

\end{document}